\documentclass[aps,prl,twocolumn,superscriptaddress, showpacs]{revtex4}
\usepackage{graphicx}
\usepackage{amsmath}
\pdfoutput=1

\newcommand{\ket}[1]{\ensuremath{\left|{#1}\right \rangle}}

\begin{document}

\title{Fast spin rotations by optically controlled geometric phases in a quantum dot}

\author{Erik D. Kim}
\author{Katherine Truex}
\author{Xiaodong Xu}
\author{Bo Sun}
\author{D. G. Steel}
 \email{dst@umich.edu}
\affiliation{The H. M. Randall Laboratory of Physics, The University of Michigan, Ann Arbor, MI 48109, USA}
\author{A. S. Bracker}
\author{D. Gammon}
\affiliation{The Naval Research Laboratory, Washington D.C. 20375, USA}
\author{L. J. Sham}
\affiliation{Department of Physics, The University of California, San Diego, La Jolla, CA, 92093-0319, USA}

\date{\today}

% We demonstrate arbitrary rotations of an electron spin qubit in a charge-tunable InAs quantum dot using a combination of circularly polarized picosecond optical pulses and spin precession about a constant external magnetic field. Arbitrary rotations are constructed from spin rotations about the $\hat{x}$ and $\hat{y}$ axes, where $\hat{z}$ is the axis of the magnetic field. In addition, we demonstrate control of the geometric phase acquired by one of the electron spin states due to cyclic $2\pi$ excitations of an optical transition in the dot. These optically induced geometric phases result in the effective rotation of the spin about the $\hat{z}$-axis and are used to demonstrate a spin phase gate.

\begin{abstract}
We demonstrate optical control of the geometric phase acquired by one of the spin states of an electron confined in a charge-tunable InAs quantum dot via cyclic $2\pi$ excitations of an optical transition in the dot. In the presence of a constant in-plane magnetic field, these optically induced geometric phases result in the effective rotation of the spin about the magnetic field axis and manifest as phase shifts in the spin quantum beat signal generated by two time-delayed circularly polarized optical pulses. The geometric phases generated in this manner more generally perform the role of a spin phase gate, proving potentially useful for quantum information applications.
\end{abstract}


\maketitle

A single charge confined in an epitaxially grown quantum dot (QD) shows considerable promise as the basic building block in a quantum computing architecture where the spin of the charge serves as the qubit \cite{LossPRA, ImamogluPRL, PierPRL}. Efforts to demonstrate the feasibility of a quantum computer based on such qubits have resulted in a number of achievements towards satisfying the DiVincenzo criteria \cite{DiViFortsch} for quantum computing. Among these achievements are spin readout \cite{ElzermanNAT, BerezSci, DannyPRL}, the demonstration of long spin coherence times \cite{Koppens, BayerSCI, CPTXD, BrunnerSCI}, spin initialization \cite{KroutvarNAT, AtatureScience, EmaryPRL, XuPRL, GerardotNAT} and the coherent control of electron spins \cite{JRPettaSCI, NowackSci, AwschSci, YamNat, BayerNatPhys}.

Of fundamental importance in executing quantum algorithms is the ability to perform a universal set of unitary operations including arbitrary single qubit operations, one of the requirements that may be met with sequential qubit rotations about two orthogonal axes \cite{BarencoPRA}. Optical approaches to performing these rotations on QD confined spin qubits are attractive as they offer the prospect of ultrafast gates using readily available laser sources \cite{ChenPRB, EmaryJPhys, ClarkPRL, EconomouPRL} and have already demonstrated fast spin rotations about the optical axis \cite{AwschSci, YamNat, BayerNatPhys}.

To obtain fast optically driven rotations about a complementary axis, Economou and Reinecke \cite{EconomouPRL} have proposed the use of geometric phases \cite{AharonovPRL} generated by cyclic $2\pi$ excitations of the optical transitions of a QD in the presence of an external DC magnetic field applied normal to the optical axis. For a properly tailored pulse, these optically induced geometric phases serve to variably alter the relative phase between the probability amplitudes of the resident spin states. If the resident spin is initially in a coherent superposition of stationary spin states, this change in the relative phase leads to an effective spin rotation about the spin quantization axis, i.e. the axis of the magnetic field. To date, however, such rotations have not been demonstrated, with recent studies \cite{YamNat, BayerNatPhys} instead relying upon combinations of pulse driven rotations about the optical axis and spin precession about the magnetic field to vary the rotation axis.

In this Report, we demonstrate the use of a narrow-bandwidth continuous-wave (CW) optical field (to simulate a long narrow-band pulse) applied between a pair of time-delayed picosecond optical pulses to \emph{optically} rotate the spin of a QD confined electron about an axis orthogonal to the optical axis $\hat{x}$. The CW field executes this rotation by driving Rabi oscillations in an optical transition in the dot involving one of the electron spin states. For each complete Rabi cycle, the optically driven spin state acquires a geometric phase $\beta$, thus altering the phase difference between the electron spin probability amplitudes. This results in a net spin rotation about the magnetic field axis $\hat{z}$, the angle depending on the properties of the optical field. Further, we show that the total geometric phase accrued to the optically driven electron spin state may be controlled by driving multiple cyclic evolutions.

\begin{figure}[t]
\includegraphics[bb = 0in 8.4in 3.25in 11in]{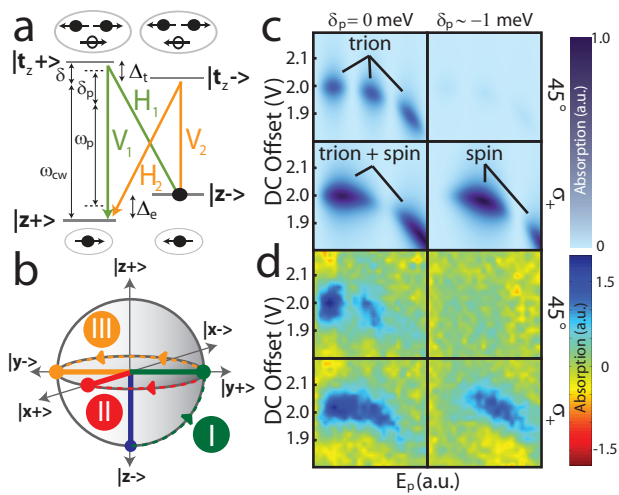}
\caption{\label{fig1} (color online). (a) Energy level diagram for a singly charged InAs QD showing the initialized state of the spin (black circle) and the properties of the pulsed and CW optical fields. The orange and green arrows illustrate the two two-photon quantum pathways from $\ket{z-}$ to $\ket{z+}$. (b) Bloch sphere for the electron spin. The blue arrow gives the orientation of the initialized electron spin vector. ``I'' shows the operation of a two-photon $\pi/2$ pulse. ``II'' and ``III'' correspond, respectively, to a $\pi/2$ rotation induced by precession and a net $\pi$ rotation resulting from an acquired geometric phase $\beta=\pi$. (c) Theoretical and (d) experimentally measured CW absorption signals in one-pulse studies for different pulse polarizations and detunings $\delta_{\text{p}}$ as a function of the sample bias DC offset and pulse amplitude $\text{E}_\text{p}$. We note that the DC offset is an order of magnitude larger than the voltage applied directly across the sample.}
\end{figure}

Experiments are performed on a single self-assembled InAs dot contained within a single-layer QD heterostructure sample that is kept at liquid helium temperatures in a magneto cryostat. $1\mu\text{m}$ diameter apertures in the Al mask on the sample surface permit optical studies of single dots. An external bias voltage provides a means of controlling both the number of electrons in a given QD and the DC Stark shift applied to the QD energy levels. In experiments, this voltage is limited to values where the dot contains a single electron. The energy level configuration for the singly-charged dot in the presence of a DC magnetic along $\hat{z}$ is shown in Fig.~\ref{fig1}(a).

% To demonstrate the use of geometric phases to rotate the electron spin, we first demonstrate the ability to optically detect arbitrary rotations of the QD confined spin.

To demonstrate the use of geometric phases to rotate the electron spin, we first demonstrate the ability to optically detect QD spin rotations about the optical and magnetic field axes. This is done by performing a series of one- and two-pulse experiments employing the CW field and mode-locked optical pulses approximately 2 ps in width. In all experiments, the spin is first prepared in a pure state [Figs.~\ref{fig1}(a),(b)] by tuning the CW optical field to the $\ket{z+}$ to $\ket{t_z+}$ transition, thereby optically pumping population from $\ket{z+}$ to $\ket{z-}$ within a few radiative cycles \cite{EmaryPRL, XuPRL}. Read-out is performed by measuring the time-averaged absorption of the same CW field on a square-law detector over a large number of repeated experiments \cite{AlenAPL, KronerPRL}, where time averaging is made possible by the fact that the repetition period of the pulses (13.2 ns) is much longer than both the spin initialization time (a few ns) and the duration of the applied spin manipulations ($< 400$ ps). These measurements effectively detect the amount of population that re-enters the $\ket{z+}$ state post initialization as a result of such manipulations.

Using this detection method, one-pulse studies are performed first to observe the optical axis rotations driven by circularly-polarized optical pulses that are detuned from the lowest-lying trion (negatively charged exciton) transitions in the dot. These pulses are then employed in two-pulse studies to observe electron spin quantum beats resulting from precession about the external magnetic field. This spin quantum beat signal forms the basis for investigations of CW-driven spin manipulations, as such manipulations manifest as modulations and phase shifts in the observed spin precession.

% optically generated geometric phase studies, as these phases lead to observable phase shifts in the spin quantum beat signal with each CW-driven trion Rabi oscillation.

In one-pulse studies, two processes can result in the generation of $\ket{z+}$ population post initialization: the excitation of trion population that then decays to the $\ket{z+}$ state and the coherent driving of two-photon Raman processes that rotate the electron spin about the optical axis. Both linearly and circularly polarized pulses can excite trion population, while only a circularly (or elliptically) polarized single pulse can drive two-photon Raman transitions. This difference in operation arises from the polarization-dependent interference of the two two-photon quantum mechanical pathways from $\ket{z-}$ to $\ket{z+}$ [Fig.~\ref{fig1}(a)]. It follows then that optically driven spin rotations may be shown by contrasting the operation of $45^\circ$ polarized $[1/\sqrt{2}(\hat{H}+\hat{V})]$ and circularly polarized $[\hat{\sigma}_+ = 1/\sqrt{2}(\hat{H}+i\hat{V})]$ pulses both on and off resonance.

Theoretical and experimental results for CW absorption studies are plotted in Figs.~\ref{fig1}(c) and (d), respectively, as a function of sample bias voltage and pulse amplitude $\text{E}_\text{p}$ for both polarizations at detunings $\delta_{\text{p}}$ of 0 meV and $\sim -1$ meV. Theoretical plots are based on numerical solutions to the density matrix equations for the four level system that include a pulse amplitude dependent red-shift of the trion transition energy attributed to pulse-generated carriers in the wetting layer. In experiments, this red shift is accounted for by adjusting the DC Stark shift at each pulse power via the sample bias.

%In experiments, the DC Stark shift of the trion transition is adjusted at each power to account for the excitation-induced shifts mentioned above.

On resonance [left column of Figs.~\ref{fig1}(c),(d)], $45^\circ$ polarized pulses show trion Rabi oscillations as a function of pulse amplitude in both theory and experiment. No spin Rabi oscillations are observed as a result of the destructive interference of the different two-photon excitation pathways. Circularly polarized pulses, however, show a combination of spin and trion Rabi oscillations, with maximum spin rotation occurring when the trion population is minimal. The strong damping seen in the experimentally observed trion Rabi oscillations for $45^\circ$-polarized pulses is not well understood at present but may be the result of off-resonant coupling of the electron to continuum states in the wetting layer \cite{GovorovPRL}. Detuning the pulses 1 meV to the red [right column of Figs.~\ref{fig1}(c),(d)] leads to highly diminished trion generation for both polarizations, resulting in a negligible absorption signal for $45^\circ$ polarization and an absorption signal due entirely to optically driven two-photon processes for circular polarization, the latter showing a complete $2\pi$ rotation of the spin. Thus, acquiring the four data sets of Fig.~\ref{fig1}(d) allows us to isolate the spin Rabi oscillation.

% Thus, a detuned circularly polarized pulse may be used to execute spin rotations about the optical axis $\hat{x}$ on timescales determined by the pulse width and with fidelities approaching unity as the detuning is increased.

% For spin rotations about an axis orthogonal to $\hat{x}$, two optically driven rotations about $\hat{x}$ with a time delay between them to allow for the free precession of the spin about the external magnetic field constitute a composite rotation about $\hat{y}$. For a pair of time-delayed, cross-circularly polarized pulses of equal pulse area ($\pi/2$), the angle of rotation about $\hat{y}$ is then given by the angle about $\hat{z}$ the spin is rotated due to spin precession. As the precession angle is time-dependent, the rotation angle about $\hat{y}$ may be controlled by varying the time-delay between the two pulses for a fixed magnetic field strength.

\begin{figure}[t]
\includegraphics[bb = 0in 9.75in 3.4in 11in]{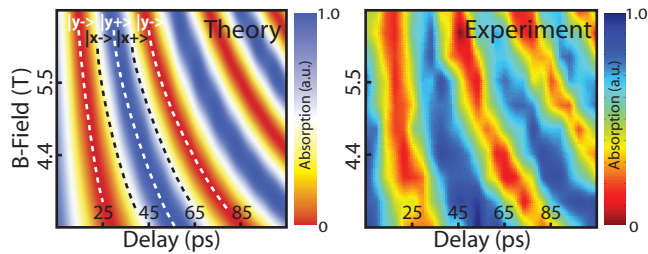}
\caption{\label{fig2} (color online). Theoretical (left) and experimental (right) absorption measurements as a function of pulse delay and magnetic field strength. Dashed lines indicate the electron spin orientation immediately before the second pulse.}
\end{figure}

For the case where the initialized electron spin is excited with a detuned circularly polarized pulse, the resulting optical axis rotation can lead to an electron spin component perpendicular to the external magnetic field. This component then precesses clock-wise about the magnetic field axis at a rate proportional to the electron Zeeman splitting energy [Fig.~\ref{fig1}(b)]. If a second, time-delayed pulse is introduced, the optical axis rotation it drives will lead to a final spin projection along $\hat{z}$ that depends on the time-delay between the pulses. As the CW read-out signal depends on the final $\hat{z}$ component of the electron spin vector, two-pulse absorption measurements will bear an oscillatory dependence on the time delay between pulses. Fig.~\ref{fig2} exhibits this oscillatory dependence in both theory and experiment for studies performed with time-delayed $\pi/2$ pulses in magnetic fields ranging from 3.3 T to 6.6 T. The orientation of the electron spin vector immediately before the second pulse indicated at selected delays. The dependence of the spin quantum beat frequency on the magnetic field strength is clearly seen, yielding an electron g-factor magnitude of $\sim 0.4$.

We now utilize this spin quantum beat signal to investigate the use of CW excitations of an optical transition in the dot to manipulate the electron spin. This scenario is realized in our case by the driving of the $\ket{z+}$ to $\ket{t_z+}$ transition by the CW field \emph{between} pulses. If a pulse incident on the initialized electron spin at time $t=0$ generates a population difference between the $\ket{t_z+}$ and $\ket{z+}$ states, the CW field will drive Rabi oscillations between these states for times shorter than the trion dephasing time while the spin precesses and is re-initialized. For trion Rabi frequencies much greater than the trion relaxation rate (yet less than the sum of the electron and heavy-hole Zeeman frequencies), each complete $2\pi$ Rabi oscillation may be considered a cyclic evolution wherein the transition wavefunction acquires an overall phase. This phase generally consists of both a dynamic component $\gamma_d$ that depends on the cycle-averaged expectation value of the Hamiltonian and the geometric component $\beta$ that may be regarded as a property of the closed curve representing the cyclic evolution in projective Hilbert space, i.e. the surface of the Bloch sphere for the $\ket{z+}$ to $\ket{t_z+}$ transition \cite{AharonovPRL, SuterPRL}. If no trion population is generated by the pulse at $t=0$, then $\gamma_d = 0$ and $\ket{z+}$ acquires a geometric phase $\beta = \pi(1 - \delta/\Omega_g)$ per complete trion Rabi oscillation where $\delta$ is the CW field detuning and $\Omega_g = \sqrt{\Omega^2 + \delta^2}$ is the generalized Rabi frequency for Rabi frequency $\Omega$ \footnote{See EPAPS Document No. for a calculation of the geometric and dynamic phases}.

In cases where the incident pulse generates a coherent superposition of electron spin states, each optically driven trion Rabi oscillation acts as an effective spin phase gate wherein the phase of one spin state ($\ket{z+}$) is altered with respect to the other ($\ket{z-}$). Further, the operation of each effective spin phase gate results in a net real space counter-clockwise rotation of the electron spin by an angle $\beta$ about the spin quantization axis $\hat{z}$ [Fig.~\ref{fig1}(b)]. These rotations manifest as net phase shifts in the spin precession signal and may be observed by performing spin quantum beat studies with higher CW powers and longer pulse delays.

\begin{figure}[t]
\includegraphics[bb = 0in 7.5in 3.0in 11in]{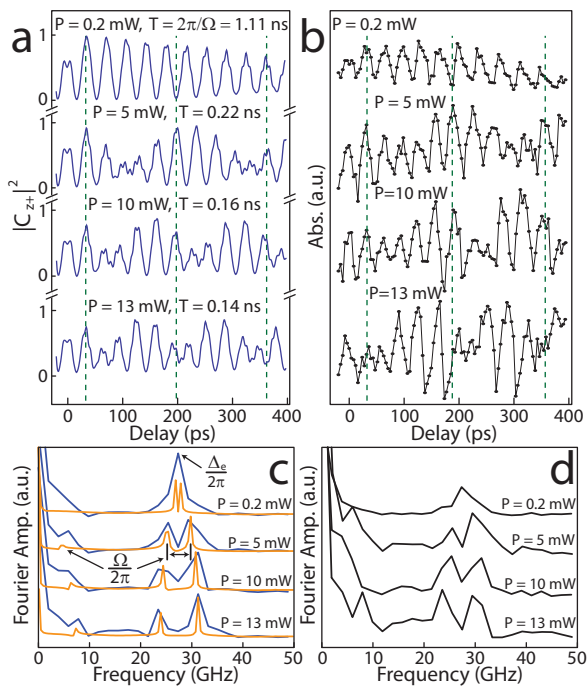}
\caption{\label{fig3} (color online). (a) Density matrix calculations of the occupation probability $|C_{z+}|^2$ immediately after the second pulse for different CW field powers and (b) the corresponding absorption measurements as a function of pulse delay, with pulses detuned 500 $\mu$eV to the red of the trion transitions. The green dashed lines serve as guides to the eye to show the change in the spin precession phase as a result of the geometric phases imparted with each complete trion Rabi oscillation. (c) Fourier spectra for calculated $|C_{z+}|^2$ traces covering 400 ps [from (a), blue curve] and 2 ns (orange curve). (d) Fourier spectrum for (b). Only a single peak at $\Delta_e/(2\pi)$ is observed in (c) and (d) for 0.2 mW scans covering 400 ps due to the limited spectral resolution provided by the delay range.}
\end{figure}

The results of such studies are given in Fig.~\ref{fig3} for a resonant CW field. Fig.~\ref{fig3}(a) plots theoretical calculations of the $\ket{z+}$ population $|C_{z+}|^2$ immediately after the second pulse for different CW powers and Fig.~\ref{fig3}(b) plots the corresponding experimental absorption measurements. Consideration of $|C_{z+}|^2$ values after the second pulse is sufficient in our case as the absorption that occurs between pulses contributes negligibly to measured signals. Immediately noticeable in both theory and experiment are the CW power-dependent modulations of the spin precession signal. These modulations are a consequence of the CW field driving population from $\ket{z+}$ to $\ket{t_z+}$ during the course of each Rabi oscillation: at times when the $\ket{z+}$ population is completely depleted, the electron spin vector points along $-\hat{z}$ and does not precess, resulting in a precession amplitude of zero. This modulation may also be seen in the analytical expression for the measured $\ket{z+}$ population for ideal $\pi/2$ pulses in the absence of decay and dephasing:
\begin{align}
|C_{z+}|^2 = \frac{1}{8}\left[3 + \cos\left(\Omega\tau\right) + 4\cos\left(\frac{\Omega \tau}{2}\right)\cos(\Delta_e \tau) \right] \label{eq1}
\end{align}
where $\tau$ is the pulse delay and the modulation arises from the third bracketed term \footnote{See EPAPS Document No. for a derivation of this expression}.

% Thus, CW driven trion Rabi oscillations are effectively ``imprinted'' on the spin precession signal and may be observed here as a function of time.

% The red circles indicate cases where the trion Rabi period is approximately equal to an integer number of spin precession periods.

In addition to spin quantum beat modulation, the effect of the optically imparted geometric phases may also be seen in Figs.~\ref{fig3}(a) and (b) when comparing the absorption signal traces for the different CW powers at the delays indicated by the green dashed lines. Since the CW field is resonant with the $\ket{z+}$ to $\ket{t_z+}$ transition, $\delta = 0$ and the geometric phase acquired by $\ket{z+}$ with each complete trion Rabi oscillation is $\pi$, manifesting as a $\pi$ phase shift in the spin precession signal. Near zero-delay (first dashed line), the spin precession phases for all traces are in phase since none have undergone a trion Rabi oscillation. Around 200 ps (second dashed line), the spin precession phases of the 5 mW and 10 mW traces are shifted by nearly $\pi$ compared to the phase of the 200 $\mu$W trace as both have undergone a complete trion Rabi oscillation. The 13 mW trace shows a phase shift of less than $\pi$ due to the fact that the delay is approaching a $\ket{z+}$ population depletion point, near which the phase changes rapidly. Around 350 ps (third dashed line) the 5 mW and 10 mW traces are once again in phase with the 200 $\mu$W trace as a result of having undergone roughly two complete trion Rabi oscillations, thus acquiring a total geometric phase of $2\pi$. This re-phasing of the 5 mW and 10 mW spin precession signals demonstrates the ability to control the total geometric phase acquired by $\ket{z+}$ by controlling the number Rabi oscillations driven by the CW field after the leading pulse.

To further support the discussion above, we plot the Fourier spectra for theoretically calculated $|C_{z+}|^2$ values and experimentally observed absorption signals at each CW power used in Figs.~\ref{fig3}(c) and (d), respectively. Theoretical spectra are provided for the calculations of Fig.~\ref{fig3}(a) (blue curve) and as well as for calculations covering 2 ns of delay (orange curve), where the latter are included to provide improved spectral resolution. The experimental spectrum is obtained from the data of Fig.~\ref{fig3}(b). Both theoretical and experimental spectra show two prominent features: a single peak at the CW Rabi frequency and a doublet centered at the spin precession frequency that increases in separation with increasing CW power. These two features may be seen as arising from the second and third terms of Eq.~\ref{eq1}. For the latter, the product of the oscillations leads to two peaks at $(\Delta_e\pm\Omega/2)/(2\pi)$ that form an Autler-Townes-like doublet, demonstrating that the CW-driven trion Rabi oscillations indeed modulate the spin quantum beat signal.

% Finally, we note that for a trion Rabi period equal to an integer number of spin precession periods, the combination of pulses and geometric phases may be used to construct purely optical spin qubit operations. The red circles in Fig.~\ref{fig3}(a) indicate cases where this condition is approximately met for the resonant CW field. Use of this condition in cases where the CW field detuning and Rabi frequency are varied would enable the execution of entirely optically driven arbitrary spin rotations. In practice, the CW laser would be gated on and off as needed for a real quantum operation.

We have demonstrated the use of CW-driven trion Rabi oscillations to control the electron spin through optically imparted geometric phases. This demonstration provides a proof of principle for the use of optical pulses to generate geometric phases on ultrafast timescales as proposed in Ref.~\cite{EconomouPRL}, which would provide a means of performing all-optical ultrafast arbitrary spin manipulations. Further, these results show that geometric phases should generally be taken into account when spin initialization or read-out is accomplished by driving optical transitions with a CW field, as such an operation itself may also induce spin rotations.

% Further, our results show that CW-generated geometric phases should be taken into account when performing optical cycling readout on QD-confined spins \cite{DannyPRL} as the readout process itself may also induce spin rotations.

This work was supported in part by ARO, NSA/LPS, ONR, NSF, AFOSR, IARPA and DARPA.

\bibliography{arxivv3}

\end{document}